\documentstyle[12pt]{article}

\newcommand{\be}{\begin{equation}}
\newcommand{\ee}{\end{equation}}
\newcommand{\bea}{\begin{eqnarray}}
\newcommand{\eea}{\end{eqnarray}}

\begin{document}
\begin{titlepage}

\begin{flushright}
{\today}
\end{flushright}
\vspace{1in}

\begin{center}
\Large
{\bf  PCT Theorem, Wightman Axioms and Conformal Bootstrap   }
\end{center}

\vspace{.2in}

\normalsize

\begin{center}
{ Jnanadeva Maharana\footnote{Adjunct Professor, NISER, Bhubaneswar}  \\
E-mail maharana$@$iopb.res.in} 
\end{center}

\normalsize

\begin{center}
 {\em Institute of Physics \\
and\\
NISER\\
Bhubaneswar - 751005, India  \\
    }

\end{center}

\vspace{.2in}

\baselineskip=24pt

\begin{abstract}
The axiomatic Wightman formulation for nonderivative conformal
field theory is adopted to derive 
conformal bootstrap equation for the four point function. 
The equivalence between PCT theorem and {\it weak
local commutativity}, due to  Jost, play a very crucial role
in axiomatic field theory.
The theorem is suitably adopted for conformal field theory
to derive the desired 
equations in CFT. We demonstrate that the two Wightman functions are analytic
continuation of each other.\\

\end{abstract}

\vspace{.5in}

\end{titlepage}

The PCT theorem is very profound. The proof of the theorem
is based on axioms of  local field theories.
It is a
fundamental probe for our basic understanding of microscopic physics.
 Pauli and L\"uder \cite{pauli,lp,grawert} presented the first proof
of the theorem. 
However, at that juncture, 
possibility of parity violation in weak interaction was not 
yet proposed by Yang and Lee
\cite{yl}. Parity nonconservation was subsequently observed experimentally. 
Thus, in the proof of  the Pauli-L\"uder 
theorem, the  violations of disctrete symmetries, such as  P, C, and T, 
in  field theories, had  not been envisaged.  
 Jost \cite{jost,pct} proved the PCT theorem from the axioms of 
 local field theories whereas earlier proofs  were based
on Lagrangian field theories. The fundamental nature of Jost's proof is
  that weak local commutativity (WLC) at the Jost points 
is necessary and 
sufficient  condition for PCT symmetry.   This aspect will be elaborated  
in the sequel.
 One of the most important
consequences of PCT theorem is that masses of particle and antiparticle
be equal. 
The best experimental test comes from the $K^0-{\bar K}^0$ mass
difference \cite{pdg}. The limit is
$-4.0\times 10^{-19}~GeV~<m_{K^0}-m_{{\bar K}^0}<4.0\times 10^{-19}~ GeV$.
Moreover, the violation of  PCT invariance of  any Wightman function
implies the violation of Lorentz invariance
  \cite{greenberg}.
Therefore, there is so much of premium on the PCT theorem.
\\

The purpose of this letter is to derive the conformal bootstrap equation
in the Wightman's formulation of axiomatic field theory \cite{pct,wight}.
The  PCT theorem is invoked to relate two four point Wightman functions. 
Our motivation is from following considerations. Let us envisage two
four point Wightman functions: 
(i) $<0|\phi(x_1)\phi(x_2)\phi(x_3)\phi(x_4)|0>$ and
(ii) $<0|\phi(x_4)\phi(x_3)\phi(x_2)\phi(x_1)|0>$ . They are also
boundary values of the analytically continued corresponding analytic
functions of complexified coordinates as will be elaborated shortly. 
As such, there seems to be no relationship between them at this juncture;
however, once PCT theorem is invoked they get related.
As often the case in study of scattering amplitudes, seemingly unrelated
amplitudes are no longer independent when we invoke a symmetry
principle. For example the reactions, $\pi N\rightarrow \pi N$, are
described by several independent amplitudes until we invoke the
principle of isotpic spin invariance of strong interaction. Then the
amplitudes $\pi^+ p\rightarrow\pi^+ p$, $\pi^- p\rightarrow\pi^- p$ and
$\pi^- p\rightarrow\pi^0 n$ are no longer independent.  
We appeal
to PCT theorem, its equivalence to weak local commutativity and
invoke the edge-of-the-wedge theorem to establish relation 
between the two Wightman functions, (i) and (ii)
defined above. It will be shown that conformal bootstrap
equation relating (i) and (ii) would be obtained. This is achieved by employing 
the conformal partial wave expansion
procedure for each of the four point functions. It is assumed in our derivation
that CPW converges for the case under consideration.   
We feel that it is a novel way to relate two different four
point functions and obtain bootstrap equation. In order to derive the
aforementioned result,   
we shall recapitulate 
 important theorems of axiomatic field theory and recall some of the 
holomorphic properties of the Wightman functions \cite{wight}.
We provide references to original papers
and to books for the benefit  of  readers.
 The derivation of the conformal bootstrap
equation is presented 
step by step sequentially starting from the important results of Wightman
formulation. The rigorous results demonstrating intimate relationships
between  analyticity, crossing and causality in CFT and
their connections with bootstrap will be presented in our  
forthcoming article \cite{jmW}.\\

The research
 activities in CFT  sprouted   
following the seminal paper of Mack and Salam \cite{ms} and a lot of
activties ensued in 1970's
\cite{fgg1,fp1,fp2,todo}. 
Migdal
introduced the idea of conformal invariance to derive bootstrap equations
for  hadronic interactions \cite{migdal}. The conformal bootstrap was
proposed by various authors in that period 
\cite{ferrara,ferrara2,polya2}.  A 
rejuvinated activity
has emerged in recent years  and the conformal bootstrap 
program has expanded
in several ditrections; these  have been reviewed recently
\cite{rev1,rev2,rev3}. 
There has been vigorous  research activities in CFT and it   
has spread in diverse directions such as the understanding of critical 
phenomena, supersymmetric conformal field theories in higher dimensions 
($D>4$). The studies of  conformal properties of gauge theories 
and gravity has drawn a lot of attention \cite{fp2}. One of the most exciting
developments is the $AdS/CFT$ correspondance conjecture of Maldacena \cite{juan}
which has strongly influenced
research in supergravity  and string theories.\\

 The conformal  transformation properties of a real scalar field,
$\phi(x)$, are
\bea
\label{coft3}
[P_{\mu},\phi(x)]=i\partial_{\mu}\phi(x),~~[M_{\mu\nu}, \phi(x)]
=i(x_{\mu}\partial_{\nu}-x_{\nu}\partial_{\mu})\phi(x)
\eea
\bea
\label{cft4}
[D,\phi(x)]=i(d+x^{\nu}\partial_{\nu})\phi(x),~~[K_{\mu},\phi(x)]
=i(x^2\partial_{\mu}-2x_{\mu}x^{\nu}\partial_{\nu}-2x_{\mu}d)\phi(x)
\eea
 $\{P_{\mu}, M_{\mu\nu} \}$ are the ten generators of the Poincar\'e
group; $D$ generates dilation and $K_{\mu}$ are the four special conformal
transformation generators; $d$ is the scale dimension of $\phi(x)$.
\\

Let us discuss some important features of conformal field theories. 
 To begin with we recall that in the perturbative approach to field theory, 
crossing property of the scattering amplitude is maintained in each order.
All Feynman diagrams, corresponding to direct channel and crossed channels, 
 are included in every order. The analyticity and unitarity
of amplitude are maintained according to stipulated rules in
perturbation theoretic  computations.  In the context of
phenomenological S-matrix theory,  which described hadronic collisions,
 crossing is assumed. The idea of bootstrap
evolved from the S-matrix philosophy. The bootstrap equations were used as
consistency conditions in the S-matrix era. The rigorous proof of crossing
for scattering amplitude was derived in subsequent years \cite{beg}. 
The bootstrap equations were introduced in conformal field theory in
order to provide a rigorous basis to the phenomenological S-matrix
notion of bootstrap. 
The structure of conformal field theories is deeply connected with the
symmetry principles. In general, a Lagrangian density is not introduced,
nor there is an action principle. Moreover, we do not invoke the
concept of asymptotic fields and interacting fields. As a consequence,
the axiomatic field theoretic techniques to compute scattering amplitude,
for example in the LSZ formulation \cite{lsz}, turn out to be inadequate. 
 Therefore, many of the rigorous results proved for scattering
amplitudes from axiomatic field theories do not hold automatically in CFT.
  Important parameters are computed in CFT to
test the theories against parameters of physical systems \cite{rev3}.\\ 

The  correlation functions in CFT are of extreme importance. Furthermore,
the analyticity and crossing properties play a crucial role in the 
study of CFT. Moreover, there is intimate relationship between causality
and analyticity. 
Therefore, analyticity, crossing and causality are three ingredients in 
the study of CFT. 
It is natural to adopt Lorenzian metric. Our choice
of the metric is
$g_{\mu\nu}={\rm diag}~(+1,-1,-1,-1)$ and we work in four dimensional
spacetime, $D=4$. We adopt Wighgtman's
axiomatic formulation to investigate aforementioned
attributes in CFT.  The importance of
Wightman function has been emphasized long ago in the intial
developmental phase of CFT \cite{fgg1,fp1,fp2,todo,polya2,luscher}. 
The axioms are \cite{wight,book1,pct,book3,book4,book5,book6}:\\
(i) There exists a Hilbert space. It is constructed \cite{fp2} with
appropriate definitions  for   CFT.\\
(ii) The theory is conformally invariant and the vacuum, $|0>$, is unique.
It  is  annihilated by all the generators of the conformal group.\\
 (iii) Spetrality:  The energy and momentum of  states
 are defined such that $0\le p^2\le\infty$ and $p_0\ge 0$. We consider a class
of theories such that the Fourier transform of
 the field, $\phi(x)$, i.e. ${\tilde \phi}(p)$,  satisfies spectrality
condition stated above
\cite{mack1,fp2}.\\
(iv) Microcausality: Two local bosonic operators commute when their separation
is
spacelike i.e. $[{\cal O}(x),{\cal O}(x')]=0$;  for  $(x-x')^2<0$. \\

The operator product expansion (OPE) of Wilson \cite{wilson} plays a crucial
role in CFT. Consider product of two real scalar field operators   ($A(x)$ is
any  real scalar field)
\bea
\label{cft1}
A(x_1)A(x_2)= \sum f_k(x_1-x_2)C_k({{x_1+x_2}\over 2})
\eea
This is a generic form of OPE. Here $\{f_k(x_1-x_2) \}$ are c-number functions 
which acquire singularities
as $(x_1-x_2)\rightarrow 0$ and operators $\{C_k \}$  are 
local in $(x_1+x_2)$.
The OPE, adopted for CFT, is very crucial.
We remark that Wilson operator
product expansion \cite{wilson} was investigated from the Wighgtman axiom
perspective by Wilson and Zimmerman \cite{wz}. Subsequently, Otterson and
Zimmermann \cite{oz} advanced those technques. \\
 
A nonderivative scalar field $\phi(x)$, satisfying $[K_{\mu},\phi(0)]=0$,
$K_{\mu}$ being the generator of special conformal transformation, 
respects Wightman axioms \cite{mack1}.  The 
Fourier transform of $\phi(x)$, ${\tilde \phi}(p)$,
 satisfies Wightman spectrality condition:
 $p\in V^+$, i.e. $p^2\ge 0$ and $p_0\ge 0$. Now on,  $\phi(x)$ stands for a 
nonderivative real conformal field and any other generic scalar field
is denoted as $A(x)$.
We need to define the the structure of the Hilbert space. 
If we consider OPE of a pair of $\phi$
fields the  form is
\bea
\label{cft2}
\phi(x_1)\phi(x_2)=\sum_n\sum_{\chi}f_n^{\chi}(x_1-x_2)
{\cal C}^{\chi}_n({{{x_1+x_2}\over 2}})
\eea
Here $f^{\chi}_n$ are c-number coefficients which encode the short
distance behaviour. ${\cal C}^{\chi}_n$ are composite fields  belonging 
to the  irreducible representations of $SU(2,2)$ which
is the covering group of the conformal group $SO(4,2)$. For the case
at hand, the set of fields  $\{{\cal C}^{\chi}_n \}$ are of nonderivative type
\cite{mack1}. It has been proved by Mack \cite{mack1} that the expansion
is convergent, assuming that it is asymptotic one.
Note that  we need infinite number
of such fields for the closure of the algebra \cite{fp2}. 
The states are constructed by appealing to 
 state$\leftrightarrow$operator  correspondences. Thus, we can 
identify state vectors of the underlying Hilbert space, ${\cal H}$. In
view of preceding remarks,  ${\cal H}$ decomposes into a direct sum of 
subspaces whose vectors belong to
irreducible representations of $SU(2,2)$.
\bea
\label{cft6}
{\cal H}=\oplus {\cal H}^{\chi}
\eea
where $\chi$ collectively stands for all the quantum numbers that
characterize an irreducible representation such as scale dimension,
Lorentz spin etc. Therefore, two normalised vectors 
$|\chi_i>\in {\cal H}^{\chi_i}$ and 
$|\chi_j>\in{\cal H}^{\chi_j}$ 
 satisfy
$<\chi_i|\chi_j>=\delta_{ij}$.
\\  

We recapitulate
the essential properties of the n-point Wightman functions, 
$W_n(x_1,x_2,...x_n)$,  defined to be  
\bea
\label{cft7}
W_n(x_1,x_2,...x_n)=<0| \phi(x_1)\phi(x_2)...\phi(x_n)|0>
\eea
in order to eventually derive the bootstrap equations. 
Note that $W_n(x_1,x_2,...x_n)$ are not ordinary functions but are
distributions. They are  defined as 
\bea
\label{cft8}
W_n[f]=\int d^4 x_1..d^4x_n W(x_1,x_2,...x_n)f(x_1,x_2...x_n)
\eea
These are linear functionals; consequently, a complex number is assigned
with the introduction of $\{ f(x_1,x_2..x_n) \}$, the Schwarzian-type functions.
$ f(x_1,x_2..x_n)$ is infinitely differentiable function
  with desired support properties in the spacetime
manifold. Thus operators of the form $\phi[f]=\int d^4x \phi(x)f(x)$
are well defined. The n-point Wightman function is
a distribution in the light of preceding remarks. Thus whenever we allude to 
the properties of $W_n(x_1,x_2,...x_n)$, it is to be kept in mind
that they are distributions; therefore, statements like convergence, limits
etc. are to be understood in this context. 
It follows from
translational invariance  that $W_n$ depends on difference
of coordinates:
$ W_n(x_1,x_2,....x_n)=W_n(y_1,y_2....y_{n-1})$, where $y_j=y_j-y_{j+1}$.
Furthermore, $W_n(\{y_j \})$ are invariant under 
Lorentz transformations: 
$W_n(y_1,...y_{n-1})= 
W_n(\Lambda_r y_1,...\Lambda_r y_{n-1})$ where $\Lambda_r$ is a real proper 
Lorentz transformation; ${\rm det}~\Lambda_r=1$. 
That physical momentum states are defined for light-like momenta i.e. 
$p^2\ge 0$ and $p_0\ge 0$, implies that the Fourier transforms of 
$W_n(\{y_j \})$, ${\widetilde W}(p_1,...p_{n-1})=0,~ {\rm unless}~ 
\{ p_j \}\in V^+$.
 Define 
complex valued function  ${\cal W}_n(\{\xi_j \}), j=1,2,..n-1$. 
 These  complex variables are  defined as 
$\xi_j^{\mu}=y_j^{\mu}-i\eta_j^{\mu}$; with the
restrictions on the real set  
 $(\{y_j,\eta_j \}),~{\rm such ~ that} ~\eta_j^{\mu}\in V^+,~{\rm and }
~-\infty<y_i^{\mu}<+\infty$;  
 defining {\it   a forward tube}, $T_{n-1}$. 
The distributions, $W_n(\{y_j \})$,  are
boundary values of the analytic functions
\bea
\label{cft9}
W_n(y_1,y_2...y_{n-1})=
{\rm lim}_{\{\eta_j \}\rightarrow 0}{\cal W}_n(\xi_1,\xi_2,...\xi_{n-1})
\eea
Notice that ${\cal W}_n(\{\xi_j \})$ are invariant under 
real Lorentz transformations.
The points, $\{\xi_j \}\in T_{n-1} $,  generate a new set of points
$\Lambda\xi_1,\Lambda\xi_2,...\Lambda\xi_{n-1}$ under arbitrary complex
Lorentz transformations where $\Lambda\in SL_+(2{\bf C})$, 
${\rm det}~\Lambda=1$. Thus the operation of $\Lambda$ on points of $T_{n-1}$ 
generates 
 a new set of points which defines 
 the extended tube $T_{n-1}'$. 
Furthermore, the complex valued analytic 
function, $W_n(\{\xi \})$,  is invariant under $SL_+(2{\bf C})$ 
and possesses a single values continuation to $T_{n-1}'$ \cite{hw}.
Note the 
{\it important difference between points lying in $T_{n-1}$ and those 
lying in $T_{n-1}'$}: 
the real points $\{y_j \}$ do not belong to the tube $T_{n-1}$ 
 whereas $T_{n-1}'$ contains 
 the real points $\{y_j \}$. 
 Where do these real points lie? Jost \cite{jost} 
proved an  important theorem.
 The Jost points are spacetime points in which
all convex combinations  of successive differences are spacelike.
For  ${\cal W}_n(\xi_1,\xi_2,....\xi_{n-1})$
 a Jost point is an ordered set $(x_1,x_2...,x_n)$.{\bf\it The Jost theorem
states} \cite{jost}: {\it A real point of} $\{\xi_1,\xi_2,..\xi_{n-1} \}$ 
{\it
lies in the
extended tube}, $T_{n-1}'$, 
{\it if and only if all real four vectors of the form}
$\sum_1^{n-1}\lambda_j\xi_j^{\mu},~\lambda_j\ge 0,~\sum_1^{n-1}\lambda_j>0$ 
{\it are
spacelike}  i.e. $(\sum_1^{n-1}\lambda_j\xi_j^{\mu})^2<0,~\lambda_j\ge 0,~
\sum_1^{n-1}\lambda_j>0$. The
 necessary and sufficient condition is that all the real points
of $T_{n-1}'$ are spacelike.\\

Recently, there have been considerable activities to investigate
analyticity properties in CFT with Lorentzian signature metric
\cite{kz,hkt,hjk,chp,sch,bg,mg}. We recall that there exists a close
relationship between analyticity and crossing.
Let us consider a four point Wightman function:
$W_4(x_1,x_2,x_3,x_4)=<0|\phi(x_1)\phi(x_2)\phi(x_3)\phi(x_4)|0>$ and
the three permutated ones. They are equal
\bea
\label{cft2}
<0|\phi(x_1)\phi(x_2)\phi(x_3)\phi(x_4)|0>=&&
<0|\phi(x_1)\phi(x_3)\phi(x_2)\phi(x_4)|0>\nonumber\\&&
=<0|\phi(x_1)\phi(x_2)\phi(x_4)\phi(x_3)|0>
\eea
 when $(x_2-x_3)^2<0$ and $(x_3-x_4)^2<0$ since $[\phi(x_2),\phi_(x_3)]=0$
and $[\phi(x_3),\phi(x_4)]=0$ due to microcausality. It should be borne
in mind that the three Wightman functions are boundary values
of analytic function, as we discuss later, and they coincide in the domain
mentioned above.
The three Wightman functions
$W_4(x_1,x_2,x_3,x_4)$, $ W_4(x_1,x_3,x_2,x_4)$ and
$W_4(x_1,x_2,x_4,x_3)$   would be analytically continued to
their corresponding tubes and extended tubes. When we look at their
Fourier transforms, the support properties will be identified. 
We have proposed a method of analytic completion when a pair of
permuted Wightman functions are considered at a time. 
In this light
we have investigated crossing and analyticity of
three point and four point functions in the Wightman formulation
of CFT \cite{mpla,jmW}.
In view of above remarks we should exercise caution 
while discussing analyticity and hence the crossing.
We proceed to present  how  the conformal bootstrap equation follow from the
above considerations.\\

 Consider the first pair of correlators, (\ref{cft2}), 
(i.e. where location of
$\phi(x_2)\leftrightarrow\phi(x_3)$ are interchanged). If we adopt the
conformal partial wave technique (CPW) 
\cite{ferrara3,polya2,rev1,rev2,rev3},
on each side of the equation we obtain the desired bootstrap equation when
$(x_2-x_3)^2<0$. We recall that, as always, one has to identify
the domain where OPE converges in order to implement  CPW expansion.
 Note, however, that it is desirable to prove that 
the two permuted
Wightman functions are
boundary values of analytic functions which coincide for spacelike separations
of real coordinates.
Thus the goal is to identify the domain of holomorphies of the two
analytic functions. This is precisely accomplished in the proof of dispersion
relations in axiomatic QFT \cite{book1}.
The s and u channel absorptive parts coincide
in a spacelike separated region and then one proves that they are analytic
continuation of each other \cite{book1}. We shall accomplish the task of
analytic continuation presently.
 This
proof of analyticity, to our knowledge is not comprehensively
investigated for conformal
field theories. We appeal to PCT theorem and its equivalence with WLC
\cite{jost,dyson58} in order to derive the bootstrap condition in a domain of
holomorphy as has been investigated in  \cite{jmW}.\\

\noindent 
{\bf\it The PCT theorem in CFT} : $W_n(x_1,x_2...x_n)$, under PCT,
 transforms as 
\bea
\label{cft10}
W_n(\phi(x_1),\phi(x_2)...\phi(x_n)) \rightarrow 
W_n(\phi(-x_n),\phi(-x_{n-1}),...\phi(-x_1))
\eea
The PCT invariance of the theory
implies
\bea
\label{cft11}
<0|\phi(x_1)\phi(x_2)...\phi(x_n)|0>=
<0|\phi(-x_n)\phi(x_{n-1})...\phi(-x_1)|0>  
\eea
 If the PCT theorem holds then for every $x_1,x_2,...x_n$ with 
each $y_j=x_j-x_{j+1}$, 
 a Jost point; the WLC condition implies
\bea
\label{cft12}
<0|\phi(x_1)\phi(x_2)....\phi(x_n)|0>=<0| \phi(x_n)\phi(x_{n-1})....
\phi(x_1)|0>
\eea
is satisfied.
\\

 We focus  only on the four point function. There is a converse statement
to Jost's theorem: if WLC  holds in a real neighbourhood of (\ref{cft12}), a 
Jost point, then
the PCT condition (\ref{cft11}) is valid everywhere.  Note that WLC implies
validity of PCT symmetry for the conformal scalar.  
  We go 
through the  following
essential steps and refer to \cite{jmW} for detailed expositions. 
The WLC theorem of Jost will be employed for  
 $W_4(x_1,x_2,x_3,x_4)$ in what follows.  
  As a consequence, of the WLC  
\bea
\label{cft13}
W_4(x_1,x_2,x_3,x_4)=W_4(x_4,x_3,x_2,x_1)
\eea
The  steps:\\
{\bf\it Step 1}. {\it Assume} that CPT theorem is valid for the conformal
theory. 
Recall that ${\cal W}_4(\xi_1,\xi_2,\xi_3)$ is a holomorphic function and 
(\ref{cft12}) holds, for $n=4$,  in the extended tube $T_3'$ and the
four point function is a boundary value of ${\cal W}_4(\xi_1,\xi_2,\xi_3)$,
\bea
\label{cft14}
{\rm lim}_{\{\eta_j \}\rightarrow 0}{\cal W}_4(\xi_1,\xi_2,\xi_3)=W_4(y_1,y_2,y_3)
\eea
Moreover, ${\cal W}_4(\xi_1,\xi_2,\xi_3)$ is invariant under proper 
complex 
Lorentz transformations, $SL_+(2{\bf C})$: 
$\{\xi_i \}\rightarrow{\Lambda\{\xi_i} \},~ \xi_i\in T_3'$. Choose a 
$\Lambda$ such
that the four complex vector $\xi_i^{\mu}\rightarrow -\xi_i^{\mu}, i=1,2,3$. 
Consequently,
\bea
\label{cft15}
{\cal W}_4(\xi_1,\xi_2,\xi_3)={\cal W}_4(-\xi_1,-\xi_2,-\xi_3)
\eea
{\bf\it Step 2}. Note that the $r.h.s.$ of (\ref{cft11}), for $n=4$,
 is also boundary
value of an analytic function.
\bea
\label{cft16}
{\rm lim}_{\{\eta_j \}\rightarrow 0} {\cal W}_4(\xi_3,\xi_2,\xi_1)=
W_4(y_3,y_2,y_1)=
<0|\phi(-x_4)\phi(-x_3)\phi(-x_2)\phi(-x_1)|0>
\eea
{\bf\it Step 3}. Consider the difference of two 4-point functions:
${\cal W}_4(\xi_1,\xi_2,\xi_3)-{\cal W}_4(\xi_3,\xi_2,\xi_1)$. 
This is holomorphic
in the domain $T_3'$. This difference vanishes for ${\rm Re}~\xi_i, i=1,2,3$ by 
CPT theorem (\ref{cft11}). Now appeal to the {\it edge-of-the-wedge theorem} 
\cite{bot,epstein} and we  conclude
\bea
\label{cft17}
{\cal W}_4(\xi_1,\xi_2,\xi_3)={\cal W}_4(\xi_3,\xi_2,\xi_1)
\eea
Consider the converse of this statement. Following Hall and Wightman \cite{hw},
 if (\ref{cft17}) holds good
in an arbitrary neighbourhood of $T_3'$ it also holds good in the extended tube. 
Moreover, if it is also valid for
passing into the boundary in the tube $T_3$ then we recover the condition 
of PCT
invariance (\ref{cft11}). Thus we conclude PCT invariance is 
equivalent to  WLC in CFT.
If we utilize the equations (\ref{cft15}) and (\ref{cft17}) then 
\bea
\label{cft18}
{\cal W}_4(\xi_1,\xi_2,\xi_3)={\cal W}_4(-\xi_1,-\xi_2,-\xi_3)
\eea
{\bf\it Remark}: Suppose we try to pass to the boundary in
the above equation for any set of $\{y_i \}$ in (\ref{cft18}). 
We encounter
the following problem. We shall not be able to get a relationship between 
the two
function, in the above equation at these arbitrary real 
points. The reason is  as $\xi_1,\xi_2,\xi_3$ 
approach
real points the real vectors are in $V^+$; whereas  the real vectors of
$-\xi_1,-\xi_2,-\xi_3$ would be in $V^-$. Note the important inference: 
at the real
point of holomorphy, this is the Jost point. Therefore,  we have the equation 
\bea
\label{cft19}
 W_4(\xi_1,\xi_2,\xi_3)=&&<0|\phi(x_1)\phi(x_2)\phi(x_3)\phi(x_4)|0>\nonumber\\
&&
=W(-\xi_3,-\xi_2,-\xi_1)=<0|\phi(x_4)\phi(x_3)\phi(x_2)\phi(x_1)|0>
\eea
This equation has important implication for the bootstrap equation as
we shall demonstrate presently.\\

Let us  consider 
$<0|\phi(x_1)\phi(x_2)\phi(x_3)\phi(x_4)|0>$. We assume that conformal partial
wave expansion is convergent for the case at hand.  Then employ the conformal
partial wave expansion by introducing a complete set of states, $\{|\Psi> \}$, 
between the product of two pairs of operators: 
$\phi(x_1)\phi(x_2)$ and $\phi(x_3)\phi(x_4)$. These states, $|\Psi>$, span
all the vectors of the Hilbert space, ${\cal H}$,  i.e. 
all  irreducible 
representations of the conformal group. The resulting equation is 
a familiar expression \cite{rev3}
\bea
\label{cft20}
W_4=<0|\phi(x_1)\phi(x_2)\phi(x_3)\phi(x_4)|0>={\bf\sum}_{|\Psi>}
<0|\phi(x_1)\phi(x_2)|\Psi><\Psi|\phi(x_3)\phi(x_4)|0>
\eea
Now invoke ${\rm state}\leftrightarrow{\rm operator}$ correspondence
and interprete $<0|\phi(x_1)\phi(x_2)|\Psi>$ as a three point function: 
$<0|\phi(x_1)\phi(x_2){\hat\Psi}|0>$ with the identification 
$|\Psi>={\hat \Psi}|0>$; ${\hat \Psi}$ represents 
the complete set of operator belonging to irreducible representations 
of the covering group. The second  matrix element on the $r.h.s$ of
(\ref{cft20}) becomes another
three point function where $<\Psi|=<0|{\hat{\bar \Psi}}$; ${\hat{\bar\Psi}}$
being the adjoint of ${\hat\Psi}$.             
Thus we express (\ref{cft20}) as
\bea
\label{cft21}
<0|\phi(x_1)\phi(x_2)\phi(x_3)\phi(x_4)|0>={\bf\sum}_{{\hat\Psi}}
{\bf\sum}_{\alpha\beta}
\lambda^{\alpha}_{\phi_1\phi_2}
{\cal W}^{\alpha\beta}_{\phi_1\phi_2{\hat\Psi}{\bar{\hat\Psi}}\phi_3\phi_4}
\lambda^{\beta}_{\phi_3\phi_4}
\eea
now $\lambda^{\alpha}_{\phi_1\phi_2}$ and $\lambda^{\beta}_{\phi_3\phi_4}$
 can be read off from the above equation.
 Furthermore,
${\cal W}^{\alpha\beta}_{\phi_1\phi_2{\hat\Psi}
{\bar{\hat\Psi}}\phi_3\phi_4}$ is the conformal partial waves (CPW) 
\cite{ferrara3,rev1,rev2,rev3}. The above equation is to be understood in
the sense that it holds in a domain where the CPW expansion converges. 
The four point Wightman function appearing on the $l.h.s.$
of (\ref{cft21}) is boundary value of an analytic function in $T_3'$.\\

Now focus on the CPW expansion 
\bea
\label{cft22}
<0|\phi(x_4)\phi(x_3)\phi(x_2)\phi(x_1)|0>=
{\bf\sum}_{{\hat\Psi},{\bar{\hat\Psi}}}
{\bf\sum}_{\alpha\beta}
\lambda^{\alpha}_{\phi_4\phi_3}
{\cal W}^{\alpha\beta}_{\phi_4\phi_3{\hat\Psi}{\bar{\hat\Psi}}\phi_2\phi_1}
\lambda^{\beta}_{\phi_2\phi_1}
\eea
Recall the equivalence theorem relating the PCT theorem
and WLC \cite{jost,dyson58}. The two expressions for Wightman functions
(\ref{cft21}) and (\ref{cft22}) are equal at those Jost points and
where conformal bootstrap conditions 
\cite{ferrara3,polya2, rev1,rev2,rev3} are valid.\\

It follows from our sequence of arguments 
that these two functions are analytic functions in extended tubes.
 Thus by invoking
Jost's theorem and Dyson's \cite{dyson58} proof of analyticity 
 it follows that  the two equations (\ref{cft21}) and (\ref{cft22}) are
analytic continuations of each other since (\ref{cft19}) holds at Jost points.
The bootstrap relation
\bea
\label{cft23}
{\bf\sum}_{{\hat\Psi}}
{\bf\sum}_{\alpha\beta}
\lambda^{\alpha}_{\phi_1\phi_2}
{\cal W}^{\alpha\beta}_{\phi_1\phi_2{\hat\Psi}{\bar{\hat\Psi}}\phi_3\phi_4}
\lambda^{\beta}_{\phi_3\phi_4}
=
{\bf\sum}_{{\hat\Psi},{\bar{\hat\Psi}}}
{\bf\sum}_{\alpha\beta}
\lambda^{\alpha}_{\phi_4\phi_3}
{\cal W}^{\alpha\beta}_{\phi_4\phi_3{\hat\Psi}{\bar{\hat\Psi}}\phi_2\phi_1}
\lambda^{\beta}_{\phi_2\phi_1}
\eea
is an equation involving spacetime coordinates satisfying Jost's condition.
It is quite remarkable that using the power of PCT theorem and WLC together
with the analyticity properties alluded to above, it is possible to 
demonstrate 
that bootstrap relation holds since the two four point functions are
analytic continuation of each other (by edge-of-wedge theorem).
Notice that if we permute a given Wightman function to obtain another one 
then the pair are analytic continuation of each other.
 We have
proved in the forthcoming  longer paper 
 \cite{jmW} that a pair of four point Wightman functions
are analytic in the unions of their domains of holomorphy i.e. the union of
the corresponding extended tubes, $T_3'$'s. Consider, as an example,
a permuted four point function:
$<0|\phi(x_1)\phi(x_2)\phi(x_4)\phi(x_3)|0>$. 
It has been shown by us that these two four point functions are analytic
functions in their corresponding extended tubes. Moreover, for
$(x_3-x_4)^2<0$ they coincide. Then one argues that the two are analytic
continuation of each other. The conformal bootstrap equation can be
obtained modulo identifying the domain of  convergence of the OPE involved.
\\

{\bf\it Remarks:} The  above bootstrap condition is not obtained specifically 
for a scalar conformal 
field theory. If we consider four point Wightman function for 
nonderivative conformal fields which belong to  irreducible
representation of conformal group then the above proof will go through
with appropriate modifications. Now the 
corresponding Wightman function will carry tensor indices as the fields
 would transform according to the representations of 
$SL(2{\bf C})\otimes SL(2{\bf C})$ and they will carry their  
conformal dimensions \cite{fp2}. Therefore, the four point function will have
a tensor structure inherited from the product of four field operators,  
each field belonging
to representation of $SL(2{\bf C})\otimes SL(2{\bf  C})$ in a general setting.
 Thus the $W_4$ will be decopmosed accordingly and transform covariantly
under representations of $SL(2{\bf C})\otimes SL(2{\bf C})$. 
The preceding arguments will 
essentially go
through. Consequently, the analyticity properties and bootstrap equations
will continue to hold. 
Therefore, we conclude that the two resulting  four point functions
will be analytic continuation of each other.
  \\
  
{\bf\it Conclusions}. We have rigorously derived that the conformal bootstrap
equations hold, for conformal scalar field $\phi(x)$,
 in the extended tube $T_3'$, for the 
four point function. The power of   the WLC 
 is crucial to prove this bootstrap condition. The proof
 is based on Wightman axioms for CFT, $\phi(x)$.
We have argued that conformal bootstrap conditions will also hold for four
point functions of conformal fields belonging  irreducible
representations of conformal group so long as they are of
 nonderivative type fields.
 The work is in progress to study analyticity and crossing
for n-point functions in conformal field theories in the present optics.\\
Acknowledgments: 
 I am indebted to {\it Ba-Gaga} for their love,
affections, patience and for my {\it raison $d'$\^etre}.

\end{document}